\documentclass[twocol]{ametsocV6.1}

\usepackage[uncertainty-mode=compact]{siunitx}
\usepackage{booktabs}
\usepackage{bm}
\usepackage{amsmath} 
\usepackage[capitalise]{cleveref} 
\usepackage{amssymb} 
\usepackage{tikz}
\usepackage{multirow}
\usetikzlibrary{shapes,arrows, arrows.meta, positioning, calc}

\title{Inferring Thunderstorm Occurrence from Vertical Profiles of Convection-Permitting Simulations: Physical Insights from a Physical Deep Learning Model}

\authors{Kianusch Vahid Yousefnia,\aff{a}\correspondingauthor{Kianusch Vahid Yousefnia, kianusch.vahidyousefnia@dlr.de} 
Christoph Metzl,\aff{a} 
Tobias Bölle,\aff{a} 
}

\affiliation{\aff{a}{Deutsches Zentrum für Luft- und Raumfahrt, Institut für Physik der Atmosphäre, Oberpfaffenhofen, Germany}
}

\abstract{Thunderstorms have significant social and economic impacts due to heavy precipitation, hail, lightning, and strong winds, necessitating reliable forecasts. Thunderstorm forecasts based on numerical weather prediction (NWP) often rely on single-level surrogate predictors, like convective available potential energy and convective inhibition, derived from vertical profiles of three-dimensional atmospheric variables. In this study, we develop SALAMA 1D, a deep neural network which directly infers the probability of thunderstorm occurrence from vertical profiles of ten atmospheric variables, bypassing single-level predictors. By training the model on convection-permitting NWP forecasts, we allow SALAMA 1D to flexibly identify convective patterns, with the goal of enhancing forecast accuracy. The model's architecture is physically motivated: sparse connections encourage interactions at similar height levels while keeping model size and inference times computationally efficient, whereas a shuffling mechanism prevents the model from learning non-physical patterns tied to the vertical grid. SALAMA 1D is trained over Central Europe with lightning observations as the ground truth. Comparative analysis against a baseline machine learning model that uses single-level predictors shows SALAMA 1D’s superior skill across various metrics and lead times of up to at least 11 hours. 
Moreover, expanding the archive of forecasts from which training examples are sampled improves skill, even when training set size remains constant.
Finally, a sensitivity analysis using saliency maps indicates that our model relies on physically interpretable patterns consistent with established theoretical understanding, such as ice particle content near the tropopause, cloud cover, conditional instability, and low-level moisture.
}

\begin{document}

\maketitle

%
%
%
\statement
This work aims to improve thunderstorm forecasting by applying machine learning to vertical atmospheric profiles from numerical weather prediction. We developed a model that incorporates physical considerations, resulting in more accurate yet computationally efficient predictions compared to conventional methods. Additionally, the model provides insights into how it identifies thunderstorm occurrence, fostering interpretability and trust. Our research demonstrates how to enhance the skill of machine learning systems in severe weather forecasting, which is crucial for supporting timely, informed decision-making in situations that impact public safety and the economy.
%
%

%

\section{Introduction}\label{sec:introduction}

Thunderstorms can have devastating impacts on society and the economy due to accompanying phenomena such as lightning, intense precipitation (including graupel and hail), and strong winds. 
Severe thunderstorms may lead to flash floods, which can result in significant  damage and endanger lives \citep{Alexandros2007,Piper2016}. 
Additionally, thunderstorms cause harm to crops and livestock, resulting in considerable economic losses \citep{Holle2014}. 
Moreover, thunderstorm occurrence affects aviation and logistics, causing costly delays and heightened safety risks \citep{Gerz2012,Borsky2019}. Although the global impact of climate change on thunderstorm frequency remains highly uncertain and varies regionally, studies suggest that thunderstorms will become more frequent in many European countries \citep{Diffenbaugh2013,Raedler2019,Taszarek2021}. This trend underscores the growing importance of accurate and timely forecasts of thunderstorm occurrence in the future.

Thunderstorm forecasts based on the extrapolation of remote sensing data (nowcasts) become less reliable beyond an hour \citep{Leinonen2023}; therefore, numerical weather prediction (NWP) is routinely used for the longer lead times required in many critical decision-making processes. Invoking mathematical models and current observations, NWP simulates the future atmospheric state encoded in terms of meteorological variables \citep{Bauer2015,Palmer2017}. In this study, we use a convection-permitting NWP model, which allows convection to be resolved without convection parameterizations \citep{Yano2018}.

The identification of thunderstorms in NWP model output is complicated by the fact that no single variable directly indicates thunderstorm occurrence. Instead, one simultaneously considers multiple predictors, which 
act as surrogate indicators of thunderstorm occurrence \citep{Sobash2011,Kober2012,Simon2018}. These predictors are motivated by a combination of experience, physical models, and domain knowledge. Examples include convective available potential energy (CAPE), precipitation rate, and relative humidity at \SI{700}{\hecto\pascal}. The occurrence of thunderstorms is then inferred from these surrogate fields based on the forecasters' expertise. 

Recently, machine learning (ML) techniques have become increasingly popular for this purpose, leveraging methods such as fuzzy logic \citep{Lin2012,Li2021}, random forests \citep{Herman2018,Loken2020}, and neural networks \citep{Sobash2020,Geng2021,Zhou2022,Jardines2024}, with the latter often proving more effective \citep{Herman2018,Ukkonen2019}. 
ML involves training algorithms to recognize patterns and make predictions based on labeled data, which serve as the ground truth. For thunderstorm forecasting, these labels are often obtained from observational data sources such as lightning detection networks \citep{Ukkonen2019,Geng2021} and radar imagery \citep{Gagne2017,Burke2020}. We recently introduced the feedforward neural network model SALAMA (Signature-based Approach of Identifying Lightning Activity Using Machine Learning), which infers the probability of thunderstorm occurrence from 21 NWP predictors related to thunderstorm activity, outperforming classification based only on NWP reflectivity for lead times up to at least $\SI{11}{\hour}$ \citep{Yousefnia2024}.

The predictors that have been used by ML models for inferring thunderstorm occurrence are single-level variables; i.e., they assign a single value  to each horizontal grid point on the NWP model domain. 
Most single-level predictors can be derived from the vertical profiles of three-dimensional meteorological variables. For example, CAPE is determined by the vertical profiles of pressure, temperature, and specific humidity \citep{Markowski2010}. There are additional processes and effects outside the scope of vertical profiles, e.g., orography, or solar radiation, which are possible sources of lift which forecasters consider when predicting convection initiation.
In what follows, however, we restrict ourselves to single-level predictors that are derived from vertical profiles.
Importantly, the information contained in a set of such single-level predictors is inherently present in the corresponding vertical profiles, though in a more complex, encoded form. Consequently, an ML model trained directly on vertical profiles should, at a minimum, match the performance of one trained on single-level predictors. Given that a model based on vertical profiles has greater flexibility to detect patterns in NWP data, we hypothesize that such a model would outperform one using only single-level predictors---this is the primary focus of the present study.

In this work, we introduce SALAMA 1D, a deep neural network trained on vertical profiles of three-dimensional meteorological variables from a convection-permitting NWP model. We thereby---and in contrast to the original SALAMA model, to which we refer as SALAMA 0D in this work---bypass the conventional use of single-level predictors. To our knowledge, this is the first study to apply neural networks directly to vertical profiles of convection-permitting forecasts for the purpose of predicting thunderstorm occurrence. Just like for the original model, we use data from ICON-D2-EPS \citep{Zaengl2015,Reinert2020}, a convection-permitting NWP ensemble model for Central Europe with a horizontal resolution of $\sim\SI{2}{\kilo\meter}$ and 65 vertical levels, operationally run by the German Meteorological Service (DWD). Lightning observations from the LINET detection network \citep{Betz2009} serve as the ground truth. On the other hand, processing vertical profiles instead of single-level predictors requires an adjustment of the SALAMA 0D architecture in order to  account for the increased input dimensionality. To keep model complexity in check, we are guided by physical principles in the design of SALAMA 1D. Specifically, a sparse layer reduces the number of parameters by promoting interactions within the same height levels, while a shuffling mechanism discourages the model from learning patterns related to the vertical grid structure. This shuffling also acts as a form of regularization, helping to prevent overfitting.
Our results demonstrate that SALAMA 1D achieves superior performance compared to SALAMA 0D, highlighting the advantage of training directly on vertical profiles.

A major issue of ML models concerns the interpretability of their output \citep{Flora2024, Yang2024, Dramsch2025}. In this work, we conduct a sensitivity analysis using saliency maps to explore how the model processes the input vertical profiles. As this analysis will reveal, our model rediscovers patterns that align with established theoretical understandings of thunderstorm development and occurrence. 

Our work demonstrates the potential of applying deep neural network models directly to complex, minimally processed input data, as opposed to relying on lower-dimensional, feature-engineered inputs fed into shallow ML models. Furthermore, we illustrate how incorporating physical constraints can enhance the model's robustness and computational efficiency, while also emphasizing the importance of gaining interpretability—and thus trust—through techniques such as saliency maps. These advancements are particularly crucial for enabling ML models to reliably assist in critical decision-making processes in severe weather forecasting.

The structure of this work is as follows: In \cref{sec:data}, we provide details on the NWP data used, the lightning observations, and the methodology for compiling ML training sets. \Cref{sec:methods} describes our ML model, including its architecture and training process. In \cref{sec:results}, we report our results, including the sensitivity analysis, while \cref{sec:conclusion} summarizes our work and elaborates on its implications with possible future research avenues.

\section{Data}\label{sec:data}
Our objective is to train an ML model---SALAMA 1D---for inferring the conditional probability of thunderstorm occurrence given an array $\bm{\xi}\in\mathbb{R}^{N_\text{f}\times N_\text{z}}$ of $N_\text{f}$ meteorological variables at $N_\text{z}$ height levels.
As a baseline for comparison, we will also train a separate ML model (SALAMA 0D, \cref{sec:methods}\ref{sec:baseline}) which infers the probability of thunderstorm occurrence from $N_\text{f}'=21$ single-level meteorological variables.
The ML task is known as binary classification. ML models for this task are trained with data sets of examples. In our case, an example is a tuple $(\bm{\xi}, y)$, where the label $y\in \{0, 1\}$ denotes the class to which the sample $\bm{\xi}$ belongs (class 1: thunderstorm occurrence, class 0: no thunderstorm occurrence).
We first collect an archive of NWP forecasts of $\bm{\xi}$ (\cref{sec:data}\ref{sec:nwp}) and the corresponding labels $y$ from lightning observations (\cref{sec:data}\ref{sec:lightning}). We then randomly draw examples from the archive to compile data sets for training, validation, and testing (\cref{sec:data}\ref{sec:training-sets}). 

\begin{figure}[htbp]
\centering
\includegraphics[width=\columnwidth]{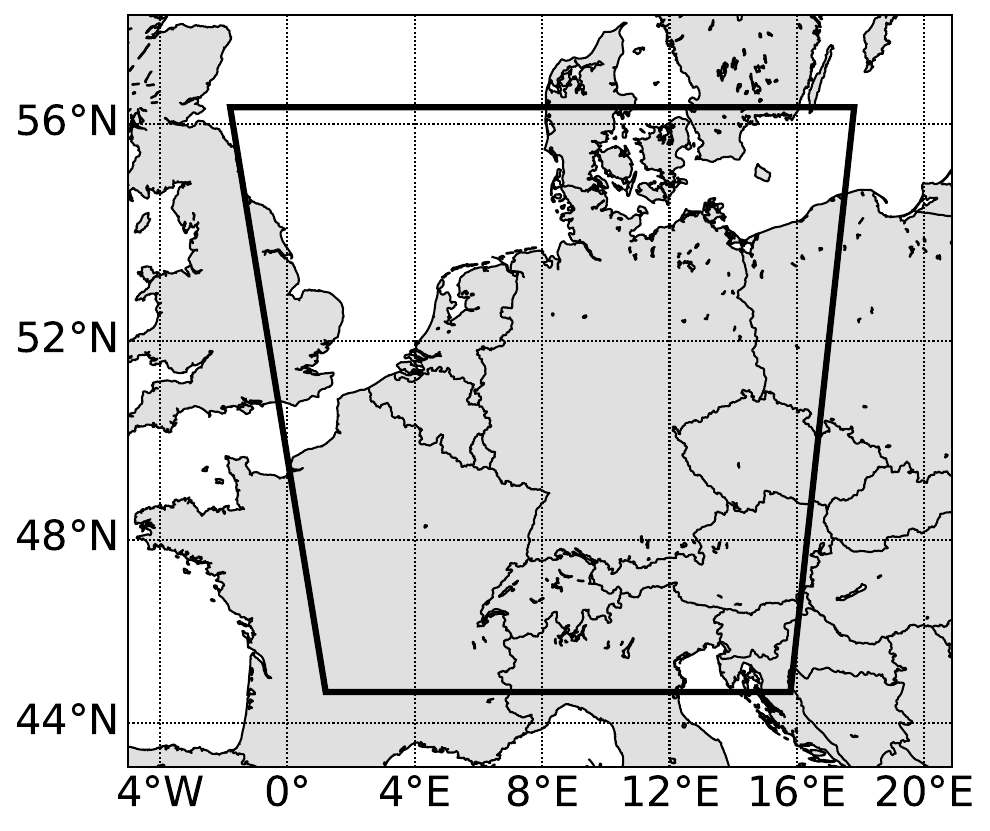}
\caption{Study region for this work, shown in a parallel projection. The polygon vertices are listed counterclockwise from the bottom-left: (\ang{44.7}N, \ang{1.2}E), (\ang{44.7}N, \ang{15.8}E), (\ang{56.3}N, \ang{17.8}E), (\ang{56.3}N, \ang{1.8}W)}
\label{fig:study_region}
\end{figure}

Data is collected for a region which encompasses Germany, as well as parts of its neighboring countries, as shown in \cref{fig:study_region}. This region roughly corresponds to the NWP model domain, which we cropped at the borders by approximately \SI{80}{\kilo\meter} to reduce boundary computation errors.

\subsection{NWP data}\label{sec:nwp}
The operational runs of ICON-D2-EPS are initialized eight times daily, starting at 0000~UTC, and produce forecasts with a time resolution of \SI{1}{\hour}.
For each full hour of the day, we collect only the latest (and, therefore, most accurate) available forecast, which is at most \SI{2}{\hour} old.
We gather forecasts for several summer months in 2021 (June, July, August), 2022 (May, June, July), and 2023 (July, August).
ICON-D2-EPS comprises 20 ensemble members, which reflect the NWP uncertainty in the initial conditions, model error, and boundary conditions \citep{Reinert2020}. We collect the forecasts for all members of the ensemble system.

For SALAMA 1D, we extract the $N_\text{f}$ variables given in \cref{tab:fields} from the forecasts; these variables correspond to the fields which are operationally available in ICON-D2-EPS on vertical levels.
We keep the fields on their native grid (vertically: $N_\text{z}=\num{65}$ non-equidistant levels, horizontally: spherical triangles) to avoid interpolation errors. For SALAMA 0D, we extract the corresponding single-level predictors \citep{Yousefnia2024}.
\begin{table}[htbp]
\caption{(Instantaneous) vertical ICON-D2-EPS field profiles used in this study.}
\label{tab:fields}
\centering
\begin{tabular}{rl}
\toprule
ICON variable & Description\\
\midrule
U & Zonal wind speed\\
V & Meridional wind speed \\
T & Temperature \\
P & Pressure \\
QV & Specific humidity \\
QC & Cloud water mixing ratio \\
QI & Cloud ice mixing ratio \\
QG & Graupel mixing ratio \\
CLC & Cloud cover \\
W & Vertical wind speed \\
\bottomrule
\end{tabular}
\end{table}

\subsection{Lightning observations}\label{sec:lightning}
In order to reconstruct the timing and location of past thunderstorm occurrences in our study region, we consult observation data from the lightning detection network LINET \citep{Betz2009}. We choose lightning observations as the ground truth for thunderstorm occurrence due to their high and uniform detection efficiency ($\geq\SI{95}{\percent}$ for LINET) and spatial accuracy (\SI{150}{\meter} for LINET), similarly to \citet{Ukkonen2019,Yousefnia2024}.

Given an NWP grid point at horizontal position $\bm{x}$ and  time $t$, we consider a thunderstorm to occur at $(\bm{x}, t)$ if a flash of lightning is detected at any $\bm{x}_\text{l}, t_\text{l}$ with
\begin{equation}
\left\| \bm{x} - \bm{x}_\text{l} \right\| < \Delta r \quad \text{and}\quad |t-t_\text{l}| < \Delta t,
\label{eq:thresholds}
\end{equation}
where $\left\| \cdot \right\|$ denotes the great-circle distance between $\bm{x}$ and $\bm{x}_\text{l}$ on a perfect sphere of the Earth with a radius of \SI{6371.229}{\kilo\meter}, as assumed in the ICON-D2-EPS model \citep{Reinert2020}.
The spatial and temporal thresholds used in this study read $\Delta r = \SI{15}{\kilo\meter}$ and $\Delta t = \SI{30}{\minute}$.

\subsection{ML data set compilation}\label{sec:training-sets}
At this stage, we have gathered a data set archive of tuples $(\bm{\xi}, y)$. Here, $\bm{\xi}$ denotes the NWP data for a particular ensemble member at a particular grid point and time, and $y$ denotes the corresponding ground truth.
We compile data sets for training, validation, and testing, by randomly drawing tuples from the archive. We provide a summary of the compiled data sets in \cref{tab:datasets} (the model configuration in the table will be introduced in this section). When compiling the data sets, there are two issues to be taken into consideration, as we will discuss next.

Firstly, to reduce correlations between the data sets, we ensure temporal separation, which is common practice \citep{Ravuri2021,Geng2021, Jardines2024}. To this end, the data of 2021 and 2022 is used for training, even days of 2023 are used for testing and odd days of 2023 are used for validation. To partition the data from 2023, we let each day begin at 0800~UTC, which we identified to be the hour of least lightning activity in our observations. The reason for shifting the start of the day is to minimize the risk of data set correlations caused by thunderstorms which persist after 0000~UTC \citep{Yousefnia2024}.

The second issue that arises is high class imbalance (observing the class ``thunderstorm occurrence'' is climatologically less likely than observing the opposite class). We estimate the relative frequency $g$ of thunderstorm occurrence expected for the test set by evaluating lightning observations for July and August from 2018 to 2022 as follows: We first assign labels to the observations by applying \cref{eq:thresholds} for each full hour and each grid point in the study region. Next, we average over the study region and each month, obtaining 10 samples of $g$ (one for each July and August of the five years above). Finally, we generate 200 bootstrap resamples of the 10-sample dataset, which yields $g=1.93^{\,+\,0.23}_{-0.24}\times 10^{-2}$ as estimate of the median and the symmetric \SI{90}{\percent} confidence interval. The problem with such a small value of $g$ is that the model might not see enough examples of the minority class during training to learn meaningful patterns. To ensure that the model is presented with sufficiently many positive examples, we compile a class-balanced training set, which contains an equal number of positive and negative examples. We arrange class balance by randomly drawing tuples from the archive and keeping a negative example only if the current number of negative examples does not exceed half of the data set target size. Thereby, we randomly undersample the majority class, as is common practice \citep{Hasanin2018,Mohammed2020,Yousefnia2024}. The validation and test set, however, are compiled in a climatologically consistent manner. The reason for this choice is to ensure that we evaluate our model in a realistic setting in which thunderstorms rarely occur. When a model trained on balanced data is used with climatologically consistent data sets, we need to calibrate the raw model output $p'$ using the following formula to obtain a well-calibrated probability $p$ \citep{Yousefnia2024}:
\begin{equation}
p = \frac{gp'}{gp' + (1-g)(1-p')}
\label{eq:calibration}
\end{equation}

Finally, in order to examine whether extending the study period from which to gather examples enhances skill, we compile two training sets for SALAMA 1D: in addition to a training set consisting of examples from 2021 and 2022 (yielding a model configuration to which we refer as SALAMA 1D-2022), we compile a second training set made up of examples from only 2021 (SALAMA 1D-2021). The models are tested, however, on the same test set.
The training set for the baseline model is equally made up of examples from only 2021, making it readily comparable to SALAMA 1D-2021. Since SALAMA 0D is trained on a different set of atmospheric variables, we cannot use the same test set as for the SALAMA 1D models. To ensure comparability, we compile the baseline model test set such that each sample corresponds to the same forecast, retrieved at the same grid point and from the same ensemble member, as its counterpart in the SALAMA 1D test set. This guarantees that any differences in model performance can be attributed to the choice of input variables.
\begin{table*}
    \caption{Summary of the data sets used for training, testing, and validation of the ML model configurations used in this study.}
    \label{tab:datasets}
    \centering
    \begin{tabular}{lccc}
    \toprule
    & SALAMA 1D-2021 & SALAMA 1D-2022 & SALAMA 0D \\
    \midrule
    \emph{Training set} & & &  \\
   Time period & Jun-Aug 2021 & Jun-Aug 2021, May-Jul 2022 & Jun-Aug 2021 \\
    No. of examples & \num{4e5} & \num{4e5} & \num{4e5}  \\
    Class imbalance & 1:1 & 1:1 & 1:1 \\
    \emph{Validation set} & & &  \\
     Time period & Jul-Aug 2023 (odd days)  & Jul-Aug 2023 (odd days) & Jul-Aug 2023 (odd days) \\
    Data set size & \num{e5} & \num{e5} & \num{e5}  \\
    Class imbalance & climat. consistent & climat. consistent & climat. consistent \\
    \emph{Test set} & & &  \\
     Time period & Jul-Aug 2023 (even days) & Jul-Aug 2023 (even days) & Jul-Aug 2023 (even days) \\
    Data set size & \num{e5} & \num{e5} & \num{e5}  \\
    Class imbalance & climat. consistent & climat. consistent & climat. consistent \\
    \bottomrule
    \end{tabular}
\end{table*}

\section{Methods}\label{sec:methods}
In this section, we present the architecture used for SALAMA 1D and give training details. In addition, we summarize the main aspects of our baseline model SALAMA 0D, which infers thunderstorm occurrence from derived single-level predictors.

\subsection{Model description}\label{sec:ML_description}
Given an input sample $\bm{\xi}$ of NWP predictors (for a given member, grid point, and forecast time), we aim to develop a model that predicts the corresponding probability of thunderstorm occurrence. Our model constitutes a lead-time-independent post-processing framework for NWP forecasts. To obtain, for example, an 8-hour forecast of the probability of thunderstorm occurrence, one needs to apply our model to an 8-hour forecast of NWP predictors. While we train on forecasts with a lead time of at most \SI{2}{\hour}, we examine in \cref{sec:results} whether the patterns learned by the ML model generalize to longer lead times.

\begin{figure*}[htbp]
\centering
\includegraphics[width=\textwidth]{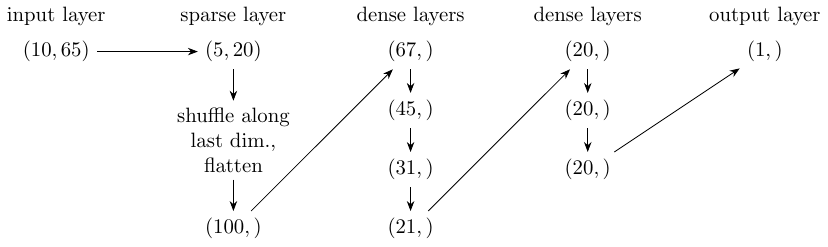}
\caption{Change in input size during a forward pass in SALAMA 1D. The sparse layer (\cref{sec:methods}\ref{sec:sparse_layer}) reduces dimensionality and shuffles the data to prevent the model from learning dependencies tied to the vertical grid structure. Additionally, the shuffling acts as a regularization technique, helping to limit overfitting. Input fields are scaled to order 1. We use rectified linear units as activation functions after the flattened sparse layer and each dense layer, and a sigmoid function to map the output layer to the open interval $(0,1)$. The sparse layer has \num{8100} trainable parameters, the other layers add \num{13226} parameters. SALAMA 1D is lightweight with a computational complexity \citep{Sovrasov2018} of roughly $22\,\text{kMAC}$ (multiply-accumulate operations). SALAMA 0D (\cref{sec:methods}\ref{sec:baseline}) requires \num{1.3}$\,\text{kMAC}$.}
\label{fig:architecture}
\end{figure*}

We use an artificial neural network model $\mathbb{R}^{N_\text{f}\times N_\text{z}} \to (0, 1)$ to describe the relationship between the input sample $\bm{\xi}$ and the corresponding probability of thunderstorm occurrence. 
The architecture of SALAMA 1D, as illustrated in \cref{fig:architecture}, combines dense layers with a sparse layer strategically designed to reduce the number of parameters. This approach addresses challenges such as overfitting and the high computational demands typically associated with large ML models. Instead of using the pruning technique \citep{LeCun1990,Frankle2019}, we incorporate physical aspects and symmetry considerations to achieve a reduction in parameters. Because translational symmetry is broken along the $z$-direction, weight sharing, as in convolutional layers, cannot be applied effectively. Instead, we implement sparse connections, allowing interactions only between field values at similar height levels (\cref{sec:methods}\ref{sec:sparse_layer}). Dense layers further downstream then construct dependencies between more distant field values. Additionally, we introduce a shuffling mechanism to ensure that the model does not rely on the vertical grid structure, forcing it to infer vertical orientation from the data itself. This design allows the model to, for instance, associate the formation of ice particles with the height of the tropopausal temperature inversion rather than a fixed height level, such as level 11. It turns out that shuffling also regularizes the model, limiting overfitting issues further.

Training is then performed analogously to \citet{Yousefnia2024}: Evaluating the training set, we scale the input fields to have zero mean and unit variance before minimizing binary cross-entropy loss via the Adam optimizer \citep{Kingma2014}. Using mini-batches of size 1000, we train for 300 epochs. After training, we inspect the validation loss as a function of epoch and select the smallest epoch for which the loss no longer decreases. \Cref{eq:calibration} is applied with the sample climatology value found in \cref{sec:data}\ref{sec:training-sets} whenever the model is used on climatologically consistent data sets. 

\subsection{Sparse connections}\label{sec:sparse_layer}
\begin{figure*}[htbp]
\includegraphics[width=0.49\textwidth]{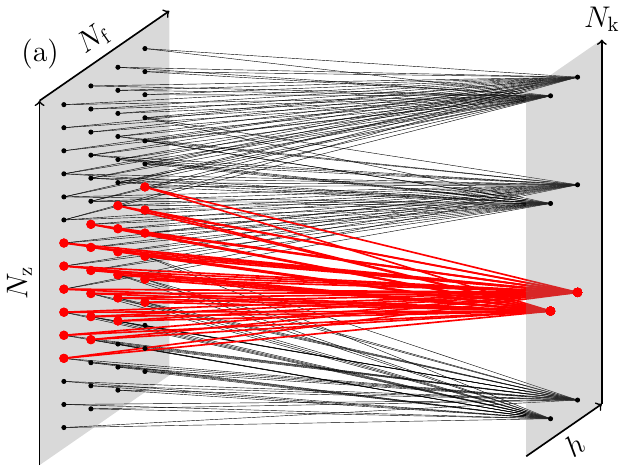}
\includegraphics[width=0.49\textwidth]{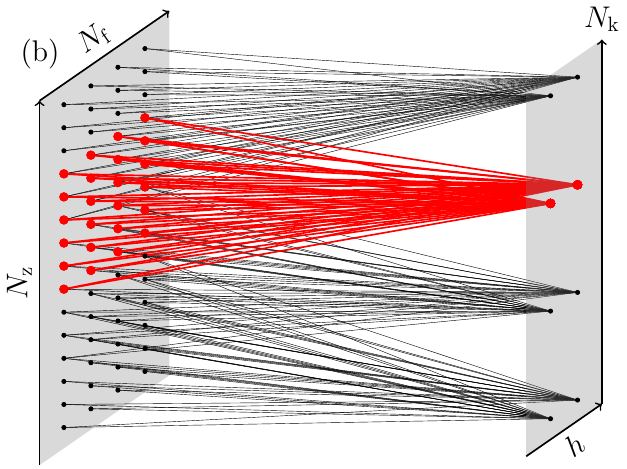}
\caption{Illustration of the connections (lines between bold dots) between the input layer of shape $(N_\text{f},N_\text{z})$ and the following layer of shape $(h, N_\text{k})$.  (a) A block of nodes of shape $(N_\text{f},k)$ in the input layer is connected to a row of $h$ nodes in the following layer. (b) Then, the block is shifted upwards by $s$ nodes and rewired with the next row of $h$ nodes of the following layer. The input layer is shown in two dimensions to help visualize each vertical field profile, but there is no spatially extended structure beyond the vertical ($z$) direction. Equally, we show the following layer in two dimensions to illustrate the yield of each group of connections in a separate row; however, this layer is flattened further downstream (\cref{fig:architecture}). For better readability, the layer sizes and hyperparameter settings used in this illustration do not correspond to those in the actual model.}
\label{fig:sparse}
\end{figure*}

In this section, we provide technical details on the sparse layer implementation, an illustration of which is provided by \cref{fig:sparse}.
The input layer is given by an array of shape $(N_\text{f}, N_\text{z})$ with a field dimension (iterating over the $N_\text{f}$ field profiles) and a height dimension (iterating over the $N_\text{z}$ vertical levels). Now, we consider a  block of shape $(N_\text{f}, k)$, where $k$ is the size of the block along the height dimension. We densely connect the nodes within this block to $h$ nodes in the following layer.
Next, we slide the block by $s$ nodes along the height dimension and, again, densely connect the corresponding nodes to $h$ subsequent nodes of the following layer. Starting with a block at the bottom of the input layer, we repeat this procedure until reaching the top of the input layer. Provided that $N_\text{z}-k$ is divisible by $s$, this procedure leads to $N_\text{k}= (N_\text{z}-k+s)/s$ blocks and produces $N_\text{k}\times h$ nodes in the following layer. We incorporate a shuffling mechanism that randomly permutes the order of the blocks for each example during training.

In contrast to a convolutional layer with a sliding kernel, all $N_\text{k}$ blocks in our sparse layer have their own set of free parameters. In total, the sparse layer contributes $N_\text{k} \times (N_\text{f}\times k +1) \times h$ parameters to the model. We have studied a large variety of $(k, s, h)$-combinations and found that skill depends barely on the particular sliding block configuration as long as a sufficiently large number of parameters is exceeded.
Setting $k = 8, s=3, h = 5$, which corresponds to the smallest model configuration with saturating skill, we obtain $N_\text{k} = 20$ blocks,  and \num{8100} trainable parameters. In comparison to a fully dense layer, the parameter size is reduced by around \SI{90}{\percent}.

\subsection{Baseline model}\label{sec:baseline}
In order to study the potential benefit of post-processing vertical NWP profiles instead of derived single-level predictors, we implement the ML model from \citet{Yousefnia2024}, which we refer to as SALAMA 0D in the following (in contrast to SALAMA 1D, which processes one-dimensional vertical profiles). SALAMA 0D uses a set of 21 established predictors related to thunderstorm activity to infer calibrated probabilities of thunderstorm occurrence.  The predictors are mapped to a single node via three dense layers with 20 nodes each. The architecture of SALAMA 0D is the result of a hyperparameter study with increasingly many layers and nodes per layer until we observed saturation in terms of validation loss \citep{Yousefnia2024}. Therefore, SALAMA 0D skill is not limited by model architecture, and any difference in skill between SALAMA 1D and SALAMA 0D can be attributed to input data differences. Furthermore, both models operate point by point, which allows for a direct and fair comparison. We train SALAMA 0D with training data from 2021 which matches the NWP model, the study region, and study period, used for SALAMA 1D (\cref{sec:data}). In contrast to the original paper, we feed the NWP data to SALAMA 0D on its native horizontal grid, reducing potential interpolation errors.

\section{Results}\label{sec:results}
We intend to investigate two aspects concerning the model skill of SALAMA 1D. Firstly, we compare SALAMA 1D to the introduced baseline model, studying the potential benefit of considering vertical profiles (instead of derived single-level predictors) for the prediction of thunderstorm occurrence. On the other hand, we are interested in examining whether extending the study period from which to gather training examples enhances skill. Therefore, we show the results for two configurations of SALAMA 1D:
\begin{itemize}
\item SALAMA 1D-2021: SALAMA 1D, trained with data from 2021.
\item SALAMA 1D-2022: SALAMA 1D, trained with data from 2021 and 2022.
\end{itemize}
The baseline model, SALAMA 0D, has been trained with data from 2021; it can, therefore, be readily compared with SALAMA 1D-2021. Note that while SALAMA 1D-2022 is trained on data from a longer study period than the two other models, the size of the training set does not change. Potential improvements in skill with respect to SALAMA 1D-2021 can, therefore, be unambiguously attributed to increased data variability. The results from the model comparison are given in \cref{sec:results}\ref{sec:model-comp}. Furthermore, we study how sensitively SALAMA 1D-2022 reacts to small input changes. The results, shown in \cref{sec:results}\ref{sec:sensitivity-analysis}, offer insight into how the model infers thunderstorm occurrence from the input.

\subsection{Model comparison}\label{sec:model-comp}

\begin{figure*}[htbp]
\centering
\includegraphics[width=\textwidth]{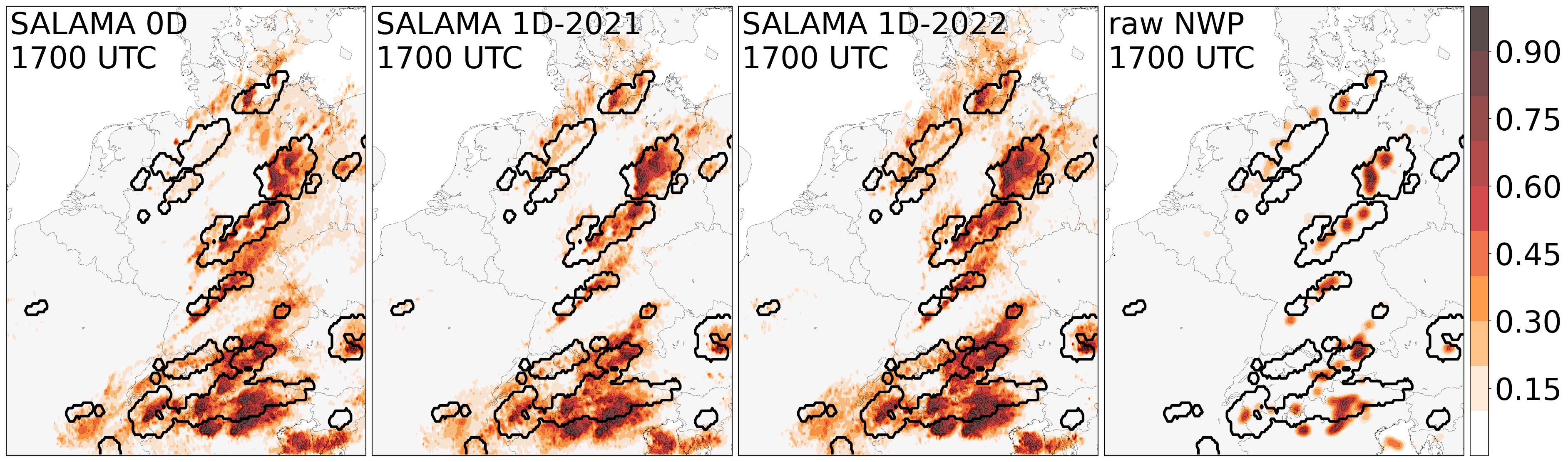}
\includegraphics[width=\textwidth]{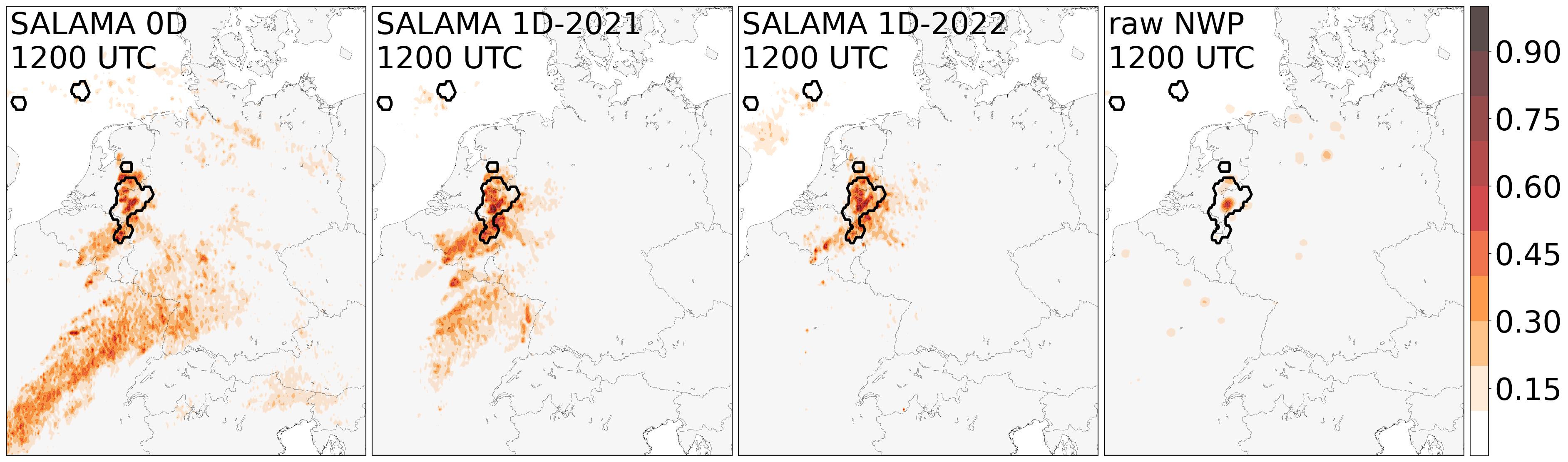}
\caption{Model probability of thunderstorm occurrence for the three models of this study, evaluated for July 24, 2023, 1700~UTC (upper panels), and August 2, 2023, 1200~UTC (lower panels). The filled contours with varying shading display the result for the
first ensemble member of ICON-D2-EPS, whereas lightning labels (\cref{sec:data}\ref{sec:lightning}) are shown as black contours. None of the dates have been used for training. NWP forecast lead times are \SI{2}{\hour} for the upper panels and \SI{0}{\hour} for the lower panels (noting that \SI{0}{\hour} forecasts have limited operational utility, as they become available only after the valid time has passed). The last column shows the probability of exceeding a reflectivity threshold of $37\,\text{dBZ}$ for the first ensemble member of ICON-D2-EPS. To this end, we compute for each pixel the fraction of pixels within a radius of \SI{15}{\kilo\meter} at which the column-maximal radar reflectivity product of ICON-D2-EPS exceeds the threshold. }
\label{fig:case_study}
\end{figure*}

In order to get a first idea of the skill of the three models, we consider two cases with thunderstorm activity in Central Europe, namely July 24, 2023, 1700~UTC (case A), and August 2, 2023, 1200~UTC (case B). These two cases were chosen since they display multiple simultaneous convective regions of varying size. In \cref{fig:case_study}, we show maps of the probability of thunderstorm occurrence for Central Europe as produced by the three models and compare them with lightning observations. The probability maps have been computed by retrieving the latest NWP forecast for each target time (case A: the 2-hour forecast of the 1500~UTC model run, case B: 0-hour forecast of the 1200~UTC run) and applying the SALAMA models to them. The ML models can be applied to all NWP ensemble members individually, producing separate probability output for each member. We show the results for only a single member. For both cases, we also show raw NWP output to see where the NWP model likely produces convection. To this end, we consider the column-maximal radar reflectivity product of ICON-D2-EPS. Specifically, for a given pixel, we compute the fraction of pixels within a radius of \SI{15}{\kilo\meter} which exceed a threshold of $37\,\text{dBZ}$ \citep[e.g.,][]{Theis2005,Roberts2008a}. Exceedance probabilities of reflectivity with thresholds between $30\,\text{dBZ}$ and $40\,\text{dBZ}$ have also been used in previous studies to identify thunderstorm occurrence  \citep[e.g.,][]{Muller2003,Leinonen2022,Ortland2023}.

Case A is characterized by intense thunderstorm activity from the Alps to Northern Germany, with roughly ten convective objects of different sizes. Most lightning contours are predicted by all three models. However, SALAMA 0D produces a significant number of false alarms. SALAMA 1D-2021 corrects many of them, especially in Southern Germany. SALAMA 1D-2022 tends to make its predictions more confidently than the other models, resulting in more contours that are filled out with high-probability pixels. On the other hand, the model seems to produce slightly more false alarms than SALAMA 1D-2021.

Thunderstorm activity in case B occurs primarily over the Benelux, while two smaller thunderstorms are observed over the North Sea. The latter two events are missed by the three models, though SALAMA 1D-2022 only misplaces the storms towards the South.
The thunderstorm over the Benelux is captured to some extent by all the models. However, the SALAMA 1D models are more confident in their predictions, producing high-probability pixels almost everywhere within the thunderstorm contour. On the other hand, they overestimate the size of the thunderstorm, resulting in false alarms directly outside the contour.  SALAMA 0D predicts a wide band of thunderstorm activity over France and Southwestern Germany, which was not confirmed by lightning observations. This region of false alarms is significantly reduced by the two SALAMA 1D models, with SALAMA 1D-2022 reducing the region to essentially zero.

The ML models, overall, align with raw NWP structures, with highest ML probability output being collocated with a high likelihood of exceeding $37\,\text{dBZ}$. On the other hand, the ML models tend to correct the areal size of simulated convection, with SALAMA 1D-2022 producing the least false alarms. Remarkably, the SALAMA 1D models can also produce high-probability output when lightning occurred but no convection has been triggered in the NWP model, as can be seen for the lightning regions over France for case A, which suggests that our ML models, SALAMA 1D-2022 in particular, may be able to correct for NWP model biases.

To compare the models quantitatively, we use the test set from \cref{sec:data}\ref{sec:training-sets}.
The first instrument which we use for comparing model skill is a reliability diagram \citep{Wilks2011,Broecker2007}. Partitioning the range $(0,1)$ of possible model probabilities into $N_\text{b}$ equidistant bins, we distribute the test examples among the bins according to the model probability they have been assigned.  For each bin $i=1,2,\ldots, N_\text{b}$, we extract the observed relative frequency $\overline{o}_i$ of thunderstorm occurrence,  the bin-averaged model probability $p_i$, and the number $N_i$ of examples per bin. A reliability diagram then consists of a calibration function and a refinement distribution. The calibration function is a plot of $\overline{o}_i$ against $p_i$, and measures whether the model probabilities are consistent with the observed relative frequency of thunderstorm occurrence, a characteristic known as reliability. A well-calibrated model exhibits a calibration function close to the 1:1 diagonal. The refinement distribution corresponds to the distribution of model probabilities. Skillful models are capable of producing well-calibrated model probabilities larger than climatology, which is referred to as resolution.

\begin{figure*}[htbp]
\centering
\includegraphics[height=0.20\textheight]{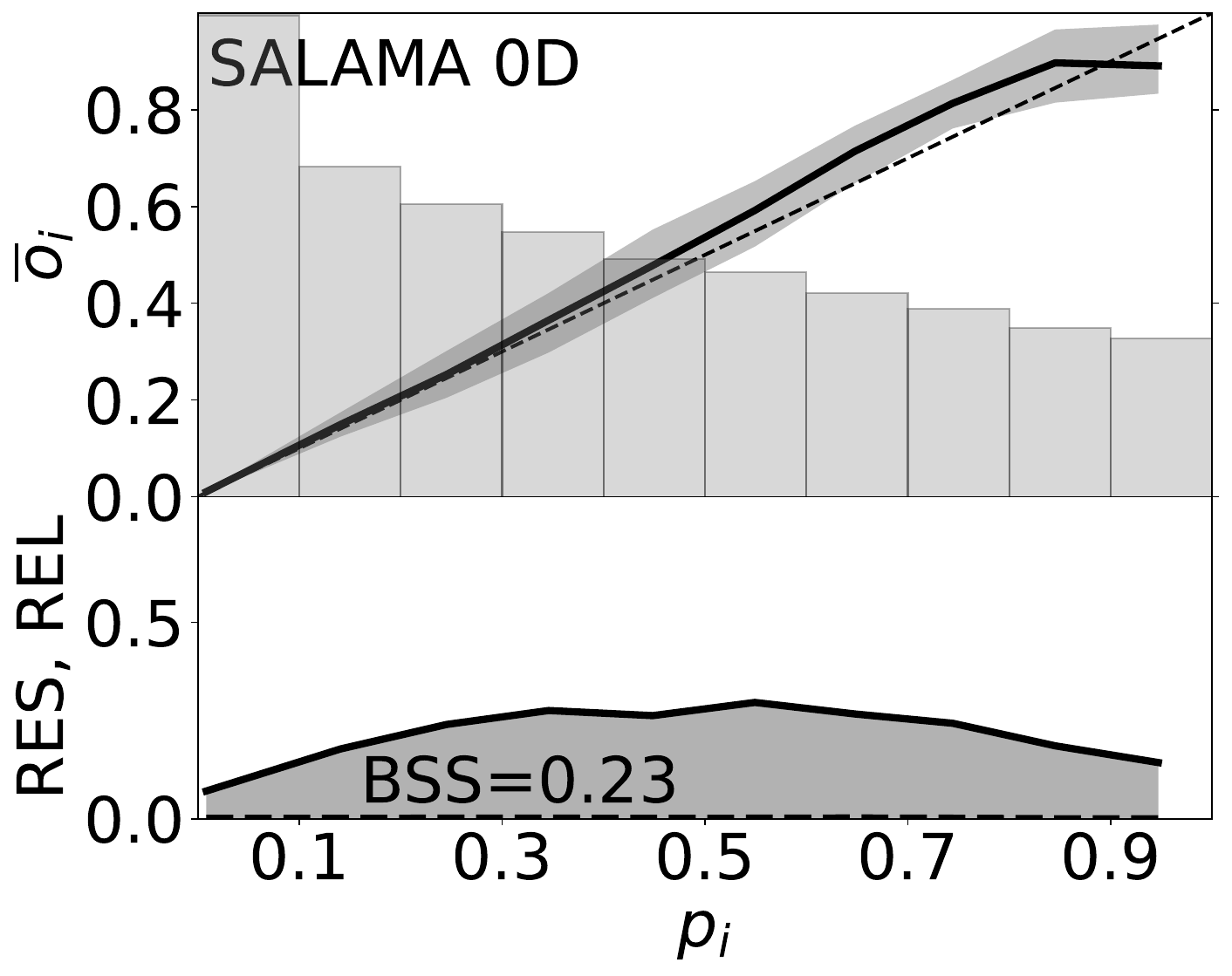}
\includegraphics[height=0.20\textheight]{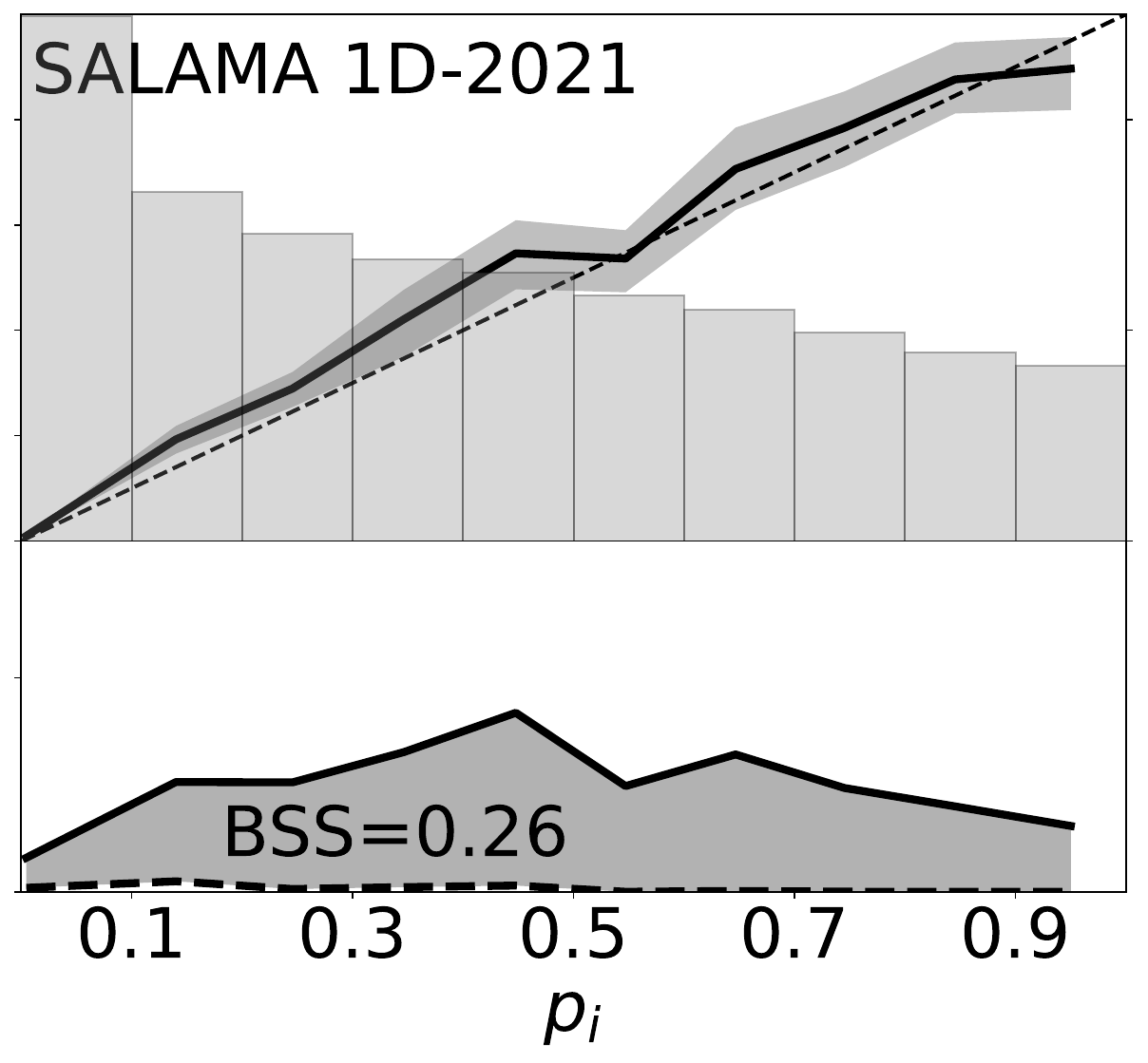}
\includegraphics[height=0.20\textheight]{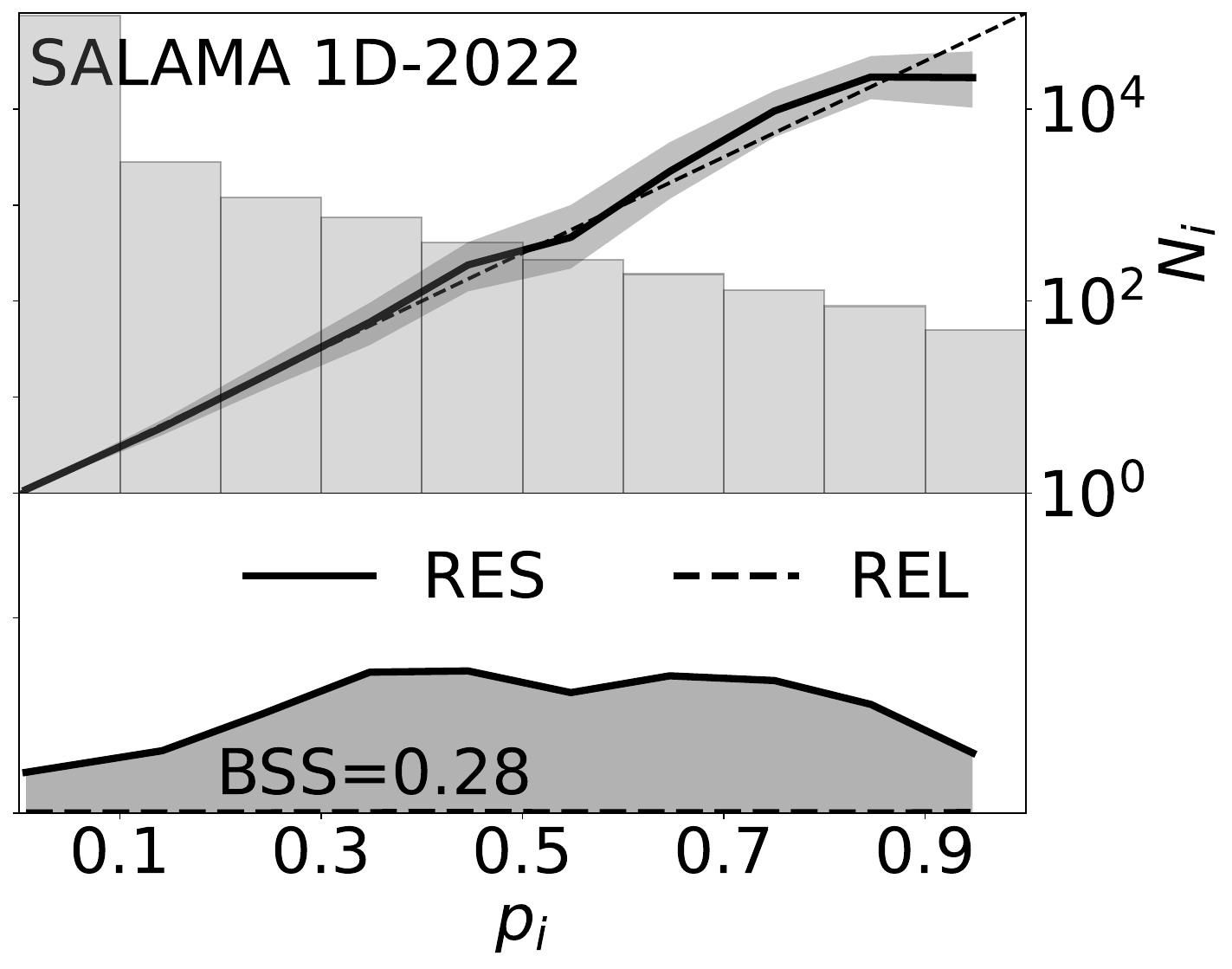}
\caption{Reliability diagram for SALAMA 0D (left), SALAMA 1D-2021 (middle), and SALAMA 1D-2022 (right). Upper panels show the calibration curve and the refinement distribution, while the lower panels display bin-wise resolution and reliability (\cref{eq:res,eq:rel}). The uncertainty on the calibration curve is obtained from \num{e4} bootstrap resamples, using day-wise block resampling, and show the symmetric \SI{90}{\percent} confidence interval. The area enclosed by bin-wise resolution and reliability corresponds to the Brier Skill Score (BSS) with climatology as reference. The BSS differences between the models are significant on a \SI{90}{\percent} confidence level, as we check in \cref{tab:skill_scores}.}
\label{fig:reliability_diagrams}
\end{figure*}

In the upper panels of \cref{fig:reliability_diagrams}, we show for each of our models the corresponding reliability diagram with $N_\text{b}=10$ bins.  All models display a similar degree of high reliability. We reiterate here that it is important to apply the analytic model calibration \eqref{eq:calibration}, otherwise high reliability could not be expected. The refinement distributions, as well, look similar. However, the model resolutions differ significantly: Following \citet{Yousefnia2024}, we introduce the bin-wise contributions to resolution and reliability
\begin{align}
\text{RES}_i &=  \frac{1/\Delta p}{g(1-g)}\frac{N_i}{N} (p_i - g)^2, \label{eq:res}\\
\text{REL}_i &=  \frac{1/\Delta p}{g(1-g)}\frac{N_i}{N} (p_i - \overline{o}_i)^2, \label{eq:rel}
\end{align}
where $\Delta p=1/N_\text{b}$ denotes bin width. The bin-wise contributions $\text{RES}_i$ to resolution are positively-oriented (``the higher, the better''), while the contributions $\text{REL}_i$ to reliability are negatively-oriented. 
In consequence, the area between $\text{RES}_i$ and $\text{REL}_i$ as a function of $p_i$ serves as a positively-oriented measure of skill. According to \citet{Murphy1973}, this area is equivalent to the Brier skill score (BSS) if a random model based on climatology (\cref{sec:data}\ref{sec:training-sets}) is applied as a reference. The lower panels of \cref{fig:reliability_diagrams} display bin-wise reliability and resolution, revealing that the increase in skill, measured by the BSS, for the two SALAMA 1D models is attributed to improved resolution. For SALAMA 1D-2021, both low- and high-probability examples enhance resolution, while for SALAMA 1D-2022, all bins with $p_i>\num{0.3}$ contribute additional improvements to resolution. 

In \cref{tab:skill_scores}, we summarize the performance of the three models using several additional skill scores established for problems with class imbalance \citep{Wilks2011,Saito2015,Yousefnia2024}. These scores are positively oriented and bounded by unity. Across all skill scores, the SALAMA 1D models consistently outperform SALAMA 0D, with SALAMA 1D-2022 showing higher skill than SALAMA 1D-2021. 
\begin{table*}[htbp]
\caption{Scores for classification skill, as defined in e.g., \citet{Yousefnia2024}, evaluated on the test set. All scores except BSS and PR-AUC require setting a decision threshold to convert probabilities to binary output. The threshold is chosen for each model such that the average fraction of examples classified as thunderstorms is equal to the observed fraction of thunderstorm examples. For this threshold, recall equals precision, and the $F_1$-score, such that only recall is reported here. Uncertainties are obtained from \num{e4} bootstrap resamples, using day-wise block resampling, and show the symmetric \SI{90}{\percent} confidence interval. The last three columns evaluate the distribution of difference in skill between the models ($\text{score(A)}-\text{score(B)}$ for model pair $\text{(A, B)}$), obtained from the bootstrap resamples, and show that all differences are significant on a \SI{90}{\percent} confidence level.}
\label{tab:skill_scores}
\begin{tabular}{p{0.2\textwidth}lll|lll}
\toprule
Skill score &  0D & 1D-21 & 1D-22  & 1D-21, 0D & 1D-22, 1D-21 & 1D-22, 0D \\
\midrule
BSS with climatology as reference & $0.234^{+\,0.037}_{-0.048}$ & $0.261^{+\,0.035}_{-0.044}$ & $0.281^{+\,0.034}_{-0.044}$ & $0.027^{+\,0.015}_{-0.013}$ & $0.020^{+\,0.008}_{-0.008}$ & $0.047^{+\,0.014}_{-0.013}$ \\
Area under the precision--recall curve (PR-AUC) & $0.397^{+\,0.055}_{-0.071}$  & $0.439^{+\,0.051}_{-0.065}$  & $0.452^{+\,0.047}_{-0.061}$ & $0.043^{+\,0.021}_{-0.019}$ & $0.012^{+\,0.011}_{-0.011}$ & $0.055^{+\,0.021}_{-0.017}$\\
Recall & $0.414^{+\,0.045}_{-0.059}$ & $0.452^{+\,0.041}_{-0.054}$ & $0.465^{+\,0.039}_{-0.050}$ & $0.038^{+\,0.017}_{-0.016}$ & $0.014^{+\,0.009}_{-0.009}$ & $0.052^{+\,0.020}_{-0.018}$\\
Critical success index (CSI) & $0.261^{+\,0.037}_{-0.045}$ & $0.292^{+\,0.035}_{-0.043}$ & $0.303^{+\,0.034}_{-0.041}$ & $0.031^{+\,0.013}_{-0.013}$ & $0.012^{+\,0.007}_{-0.007}$ & $0.043^{+\,0.015}_{-0.014}$\\
Equitable threat score (ETS) & $0.250^{+\,0.035}_{-0.043}$ & $0.281^{+\,0.033}_{-0.042}$ & $0.293^{+\,0.032}_{-0.039}$ & $0.031^{+\,0.013}_{-0.013}$ & $0.012^{+\,0.007}_{-0.007}$ & $0.043^{+\,0.015}_{-0.014}$\\
\bottomrule
\end{tabular}
\end{table*}

So far, we have worked with a test set that consists of examples from NWP forecasts with a lead time of at most \SI{2}{\hour} (\cref{sec:data}\ref{sec:nwp}). Next, we examine systematically how model skill depends on NWP forecast lead time.  For this purpose, we generate test sets in which the examples result from NWP forecasts with a fixed lead time. Just like before, these test sets with fixed lead times are sampled in a climatologically consistent manner and consist of \num{e5} examples. The target times are drawn from the same test days as in \cref{sec:data}\ref{sec:training-sets}.

\begin{figure}[htbp]
\centering
\includegraphics[width=\columnwidth]{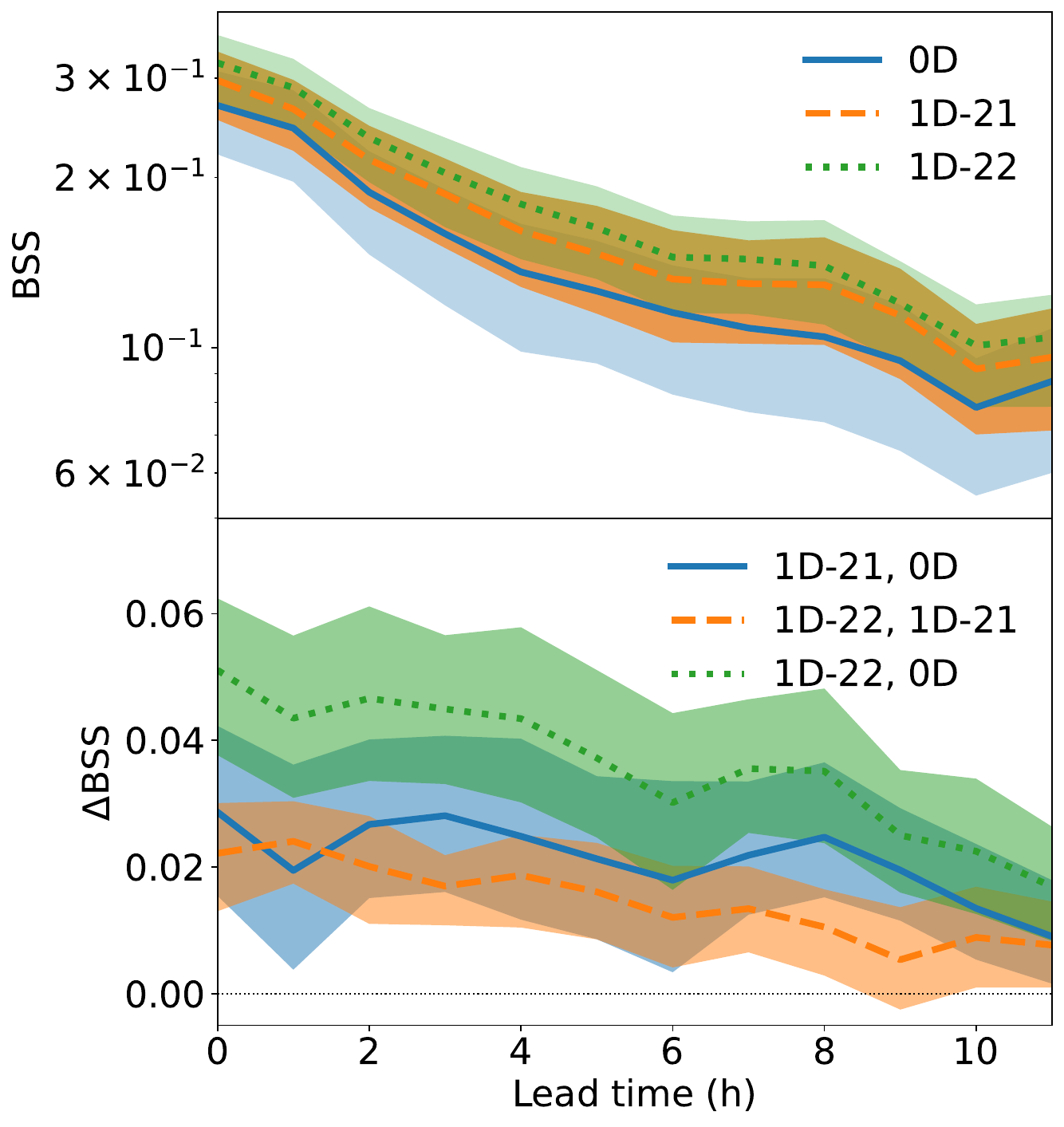}
\caption{Lead-time dependence of model skill, quantified by the BSS with climatology as reference. Upper panel shows the BSS for the individual models, while the lower panel shows the skill difference $\Delta \text{BSS} = \text{BSS(A)}-\text{BSS(B)}$ between model pair $\text{(A, B)}$. Uncertainties are obtained from \num{e4} bootstrap resamples, using day-wise block resampling, and show the symmetric \SI{90}{\percent} confidence interval. }
\label{fig:skill_leadtime}
\end{figure}

The lead-time dependence of skill is shown in the upper panel of \cref{fig:skill_leadtime}. While we measure skill in terms of the BSS, we have checked that the results of this section do not qualitatively change when considering a different skill score. All three models exhibit an approximately exponential decrease in skill. The rate at which skill decreases is very similar for the three models. This suggests that the decrease of skill is not model-specific but results from an increasing NWP forecast uncertainty, which is consistent with previous work \citep{Yousefnia2024}. As a consequence, the SALAMA 1D models' superior skill for low lead times is passed on to longer lead times. In the lower panel of \cref{fig:skill_leadtime}, we show the difference in skill as a function of lead time for all model pairs. Again, we find that the SALAMA 1D models consistently outperform SALAMA 0D at a confidence level of \SI{90}{\percent}, with SALAMA 1D-2022 showing higher skill than SALAMA 1D-2021.

It is worth noting that the decrease in skill of SALAMA 0D is stronger than reported in \citet{Yousefnia2024}. There, initial skill decreased by at most \SI{30}{\percent} after \SI{11}{\hour}, while the decrease here is approximately twice as much. This may result from using more diverse test sets in this study (we use twice as many test days to compile the training set). 

\subsection{Interpretability study}\label{sec:sensitivity-analysis}
\begin{figure*}[htbp]
\centering
\includegraphics[width=\textwidth]{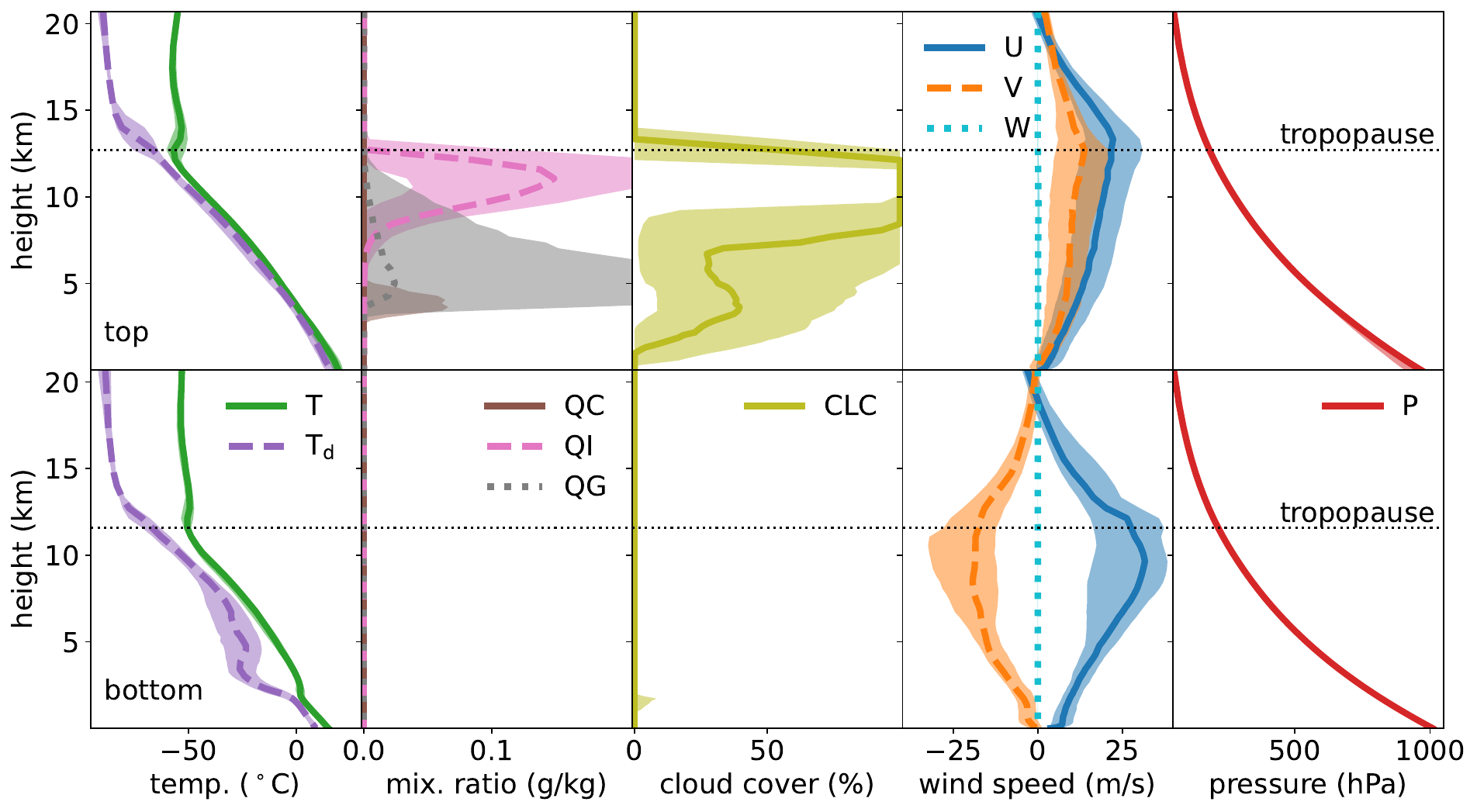}
\caption{Average vertical field profiles for the top probability percentile of test set samples (upper row) and the bottom percentile (lower row), with annotated levels of the tropopause. We convert average specific humidity to dew point temperature T$_\text{d}$ for a  comparison with T. Shaded bands correspond to the symmetric \SI{50}{\percent} confidence interval.}
\label{fig:avg_profiles}
\end{figure*}

For the remainder of this section, we focus on SALAMA 1D-2022, referring to it simply as SALAMA 1D. Our goal is to gain insight into how our model classifies input. We start by inspecting how the vertical profiles of the SALAMA 1D input fields look like on average for test set samples to which our model assigns a particularly high or low probability of thunderstorm occurrence. \Cref{fig:avg_profiles}, shows the average vertical profiles for the top and bottom probability percentile of test set samples. Shaded bands denote the symmetric \SI{50}{\percent} confidence interval. Note that we converted specific humidity to dew-point temperature $T_\text{d}$ for a more straightforward comparison with temperature. For better orientation within the panels, we also plot the average tropopause height, which we compute according to the WMO definition \cite[lowest level where the lapse rate drops to $\leq 2\,$K/km, with the average lapse rate within $2\,$km above remaining $\leq 2\,$K/km,][]{WMO1957}. 

The first column of panels in \cref{fig:avg_profiles} shows temperature and dew-point temperature $T_\text{d}$ for the two percentiles. The top percentile troposphere displays more moisture than the bottom percentile, in particular in 2-\SI{5}{\kilo\meter} height. This is consistent with the documented importance of moisture for thunderstorm development, as the buoyancy of rising air parcels is otherwise reduced by dry-air entrainment \citep{Zhang2003,Peters2023,Marquis2023}. 
In the second column in \cref{fig:avg_profiles}, we show average profiles of mixing ratios of cloud water (QC), cloud ice (QI), and graupel (QG). This suggests that the model uses non-vanishing profiles of QI and QG to discriminate between the top and the bottom percentile. For high-probability samples, ice particle content peaks close to the tropopause at a height of 10-\SI{12}{\kilo\meter}, consistent with measurements of vertical hydrometeor distributions \citep[e.g.,][]{Vivekanandan1999,Hubbert2018}. 
The third column in \cref{fig:avg_profiles} displays the average profiles of cloud cover (CLC). In general, our model associates non-vanishing CLC with a high probability of thunderstorm occurrence. In particular, close to the tropopause, CLC tends to be \SI{100}{\percent}. This is consistent with anvil cloud top levels \citep{Markowski2010}.
In the fourth column in \cref{fig:avg_profiles}, we show vertical profiles of the three wind components U, V, and W. It is noteworthy that high-probability samples tend to have southwesterly wind profiles, whereas samples from the bottom percentile display northwesterly winds. This is consistent with studies on the typical propagation direction of thunderstorms in Central Europe \citep{Hagen1999}. On the other hand, vertical profiles of W vanish for both percentiles. We presume that due to convection displacement errors in the NWP model, the updraft regions within simulated deep convection rarely match observed lightning observations. Therefore, SALAMA 1D may have learned not to rely on W for inferring thunderstorm occurrence.
The last column of \cref{fig:avg_profiles} shows the average profiles of pressure. Both profiles appear to be essentially hydrostatic. Surface pressure tends to be lower for the top percentile than for the bottom percentile.

Comparing the average profiles of the SALAMA 1D input fields for the two percentiles is useful to get a first idea whether our model separates the thunderstorm class from the majority class in a physically-interpretable manner. On the other hand, this analysis does not inform about the relative importance of the individual atmospheric variables. Therefore, we conduct a linear sensitivity analysis of the conditional probability $f$ of thunderstorm occurrence (SALAMA 1D model output) with respect to the input. The general idea is to consider for a given input sample $\bm{\xi} = (\xi_{ij}) \in \mathbb{R}^{N_\text{f}\times N_\text{z}}$ the partial derivatives of $f$:
\begin{equation}
S_{ij}(\bm{\xi}) \equiv \frac{\partial f(\bm{\xi})}{\partial \xi_{ij}}
\label{eq:saliency_signed}
\end{equation}
The term $S_{ij}(\bm{\xi})$ constitutes a measure of how much $f$ reacts to changes in $\xi_{ij}$ and, therefore, quantifies the importance of $\xi_{ij}$ to the outcome. $S_{ij}(\bm{\xi})$ is commonly referred to as saliency in the ML literature \citep{Simonyan2014, Li2022} while meteorologists might know it as adjoint sensitivity \citep{Errico1997,Warder2021}.

In order for saliency values $S_{ij}(\bm{\xi})$ to be comparable across all indices $i,j$, we need to scale the input fields appropriately. This scaling accounts for the fact that the fields have different units and vary differently from one sample to another.
Consider an unscaled input field $\tilde{\xi}(z)$ (e.g., pressure), which we take as a function of height $z$ above ground. We define the corresponding scaled fields as 
\begin{equation}
    \xi(z) = \frac{\Tilde{\xi}(z)-\mu_\xi(z)}{\sigma_\xi(z)},
\end{equation} 
where $\mu_\xi(z)$ and $\sigma_\xi(z)$ encode a characteristic background climatology for $\tilde{\xi}(z)$. We define them as
\begin{align}
\mu_\xi(z) &= P_{50}\left(\tilde{\xi}(z)\right) \\
\sigma_\xi(z) &= \frac{P_{99}\left(\tilde{\xi}(z)\right)-P_{1}\left(\tilde{\xi}(z)\right)}{2},
\end{align}
where $P_n$ stands for the $n$th percentile of the variable in the brackets, evaluated for the training set. We then compute saliency with respect to the scaled fields. 

\begin{figure}
\centering
\includegraphics[width=\columnwidth]{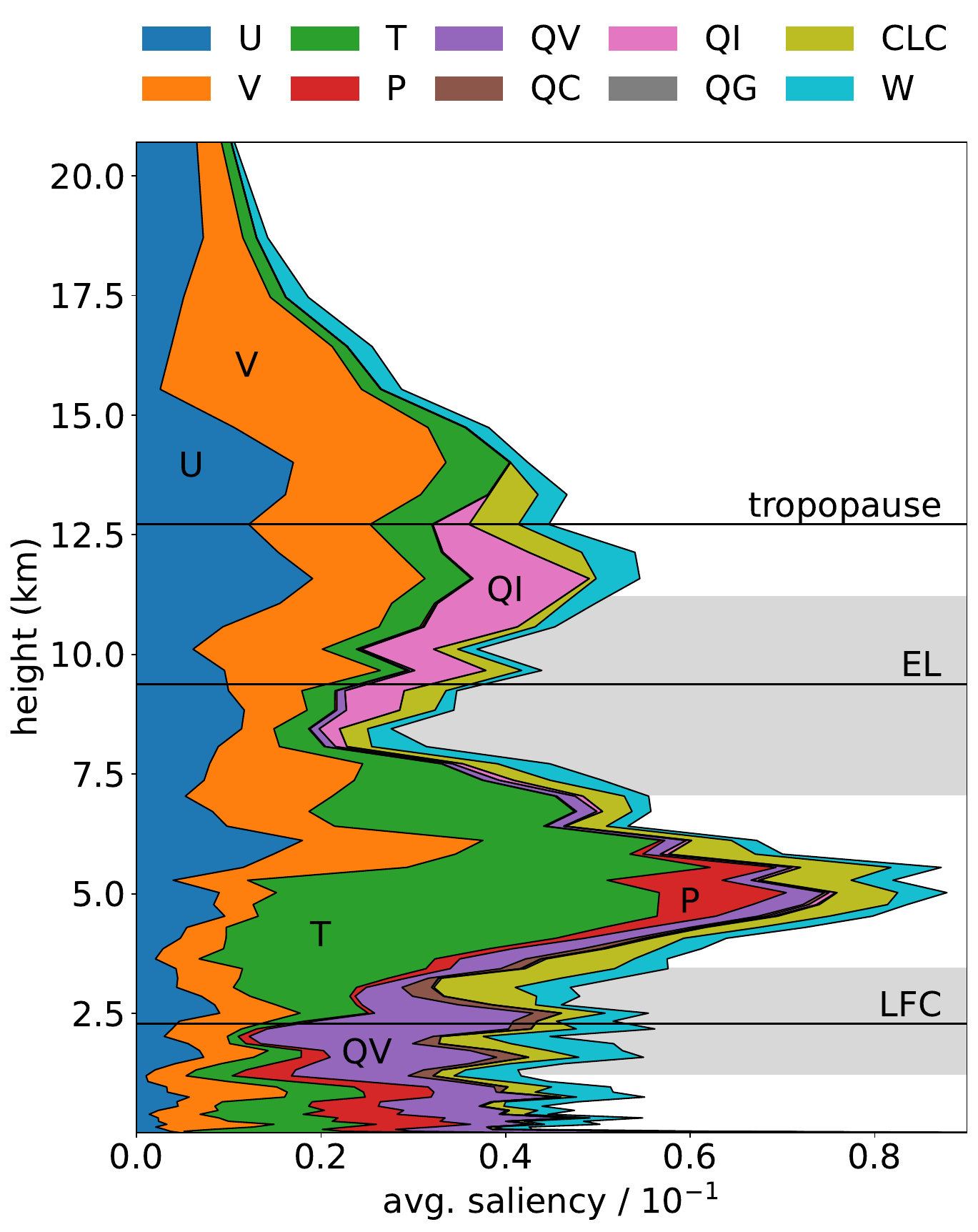}
\caption{Vertical profiles of average saliency for the top percentile of test set samples (based on model probability), with annotated levels of the tropopause, as well as the LFC and EL of a mixed-layer parcel. Saliencies for the different fields are stacked on top of each other in the order given in \cref{tab:fields}.}
\label{fig:saliency_map}
\end{figure}

While saliency varies from one sample to another and can be used to interpret individual predictions, we propose averaging over the top probability percentile to obtain more robust insight on the input fields and height levels which contribute most to high-probability output.
To this end, we compute $|\langle S_{ij}\rangle|$, where the angle brackets denote averaging over the top percentile samples. We take the absolute value (absolute saliency), as we consider feature importance to be linked to the (sample-averaged) intensity of the ML model's linear response, irrespectively of whether this response is positive or negative. In what follows, we refer to $|\langle S_{ij}\rangle|$ simply as saliency, i.e., with sample-averaging and taking the absolute value being implicitly implied unless stated otherwise. The resulting saliency maps for the two percentiles are shown in \cref{fig:saliency_map}. Note that average saliencies of the different fields are stacked on top of each other. As a consequence, the saliency envelope quantifies how much individual height levels affect the model outcome.

The saliency envelope displays two distinct peaks, at $z=\SI{12}{\kilo\meter}$ and $z=\SI{5}{\kilo\meter}$, respectively. The upper-level peak receives the most contributions by horizontal wind speed. Indeed, the saliencies of U and V are maximal near the tropopause, where the average horizontal wind velocity difference between the top and the bottom percentile is greatest (\cref{fig:avg_profiles}). This suggests that the model relies to a considerable extent on the learned climatological propagation direction of thunderstorms. Conversely, W saliency is approximately one order of magnitude smaller than the saliencies of U and V. QI saliency is maximal at the top of the troposphere, contributing significantly to the upper-level peak, as well. QI is present only at this height (\cref{fig:avg_profiles}), which suggests that the model actively takes ice particle content into consideration to infer thunderstorm occurrence. The only other hydrometeor field with significant non-vanishing saliency is specific humidity (QV). QV saliency peaks near the LFC, coinciding with the vicinity of maximal tropospheric moisture (\cref{fig:avg_profiles}). Similarly, CLC saliency is low but non-vanishing at all heights with non-zero CLC (\cref{fig:avg_profiles}).

Next, we turn to the mid-level peak. Apart from horizontal wind velocity, the mid-level peak receives most contributions from temperature. This may be partly due to convection feeding back to the temperature field. However, since pressure saliency is non-vanishing only at the mid-level peak and near the surface, we hypothesize that the ML model also reconstructs lapse rates. 
To understand this, note that our model is not informed about the height of individual vertical levels; rather, these levels are randomly shuffled (\cref{sec:methods}\ref{sec:ML_description}). Thus, our model can reliably infer heights levels---and in particular level spacings---only from pressure, which monotonously decreases with height (\cref{fig:avg_profiles}). Therefore, we expect pressure saliency to be a proxy for how much the model relies on vertical gradients. Finally, as pressure saliency contributes to the mid-level peak and temperature saliency is high, we conjecture that our model reconstructs mid-level lapse rates. This is supported by the fact that the mid-level peak is bounded by the LFC and the EL of parcels lifted from the mixed layer, meaning that such parcels are buoyant for height levels in the vicinity of the mid-level peak. Positive buoyancy occurring in a conditionally unstable troposphere is known to be a crucial ingredient for thunderstorm development \citep{Doswell1996} and constitutes the basis for several traditional thunderstorm predictors, such as convective available potential energy (CAPE). 

\begin{figure*}
    \centering
    \includegraphics[width=\textwidth]{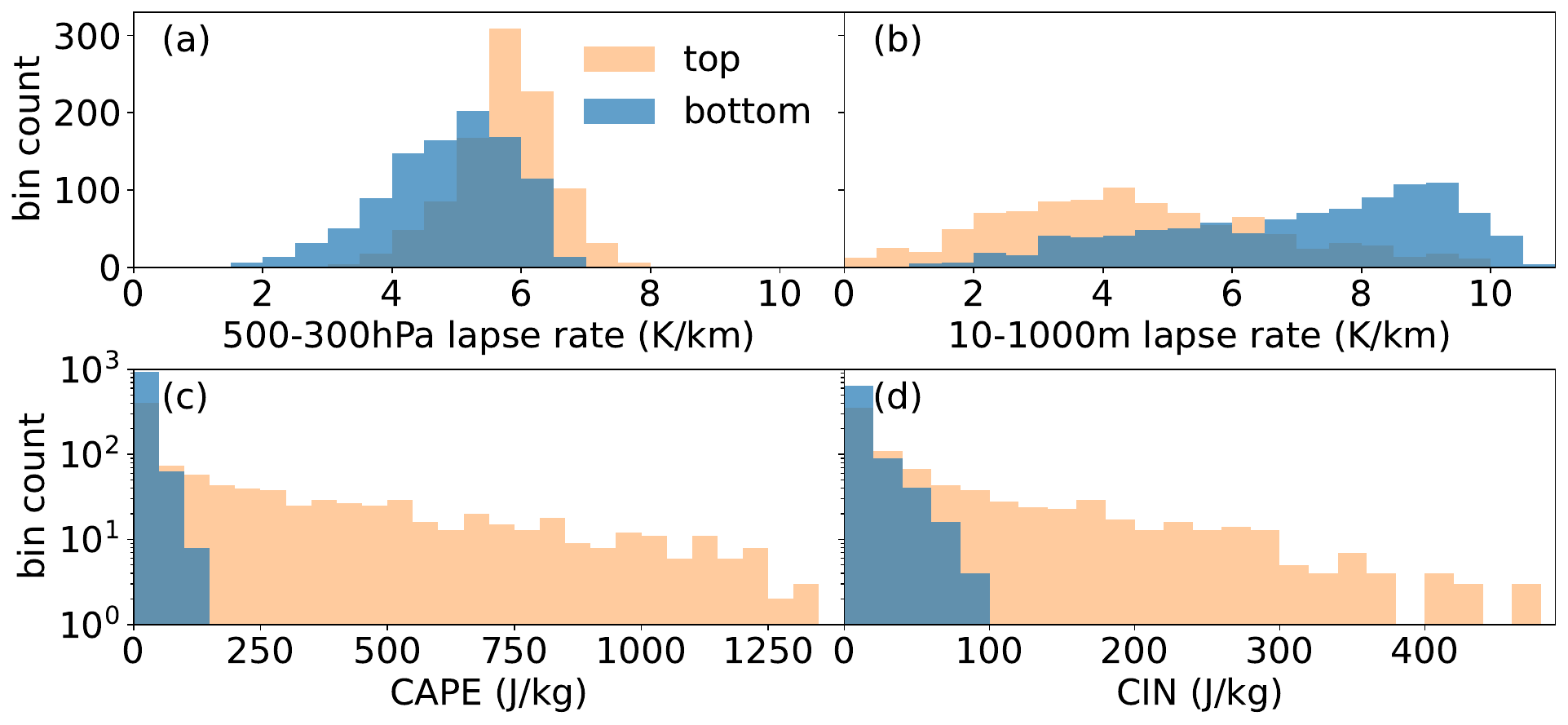}
    \caption{Distribution of mid-level lapse rates (a), near-surface lapse rates (b), mixed-layer CAPE (c) and mixed-layer CIN (d) for the top and bottom probability percentile of test set samples.}
    \label{fig:lapse-rates}
\end{figure*}

To test whether SALAMA 1D considers mid-level lapse rates, we show in \cref{fig:lapse-rates}a the distribution of 500-\SI{300}{\hecto\pascal} lapse rates for the top and bottom probability percentile. Indeed, essentially all high-probability samples are associated with conditionally unstable mid-level lapse rates, whereas the bottom percentile distribution extends further into absolutely stable mid-level lapse rates. 
On the other hand, the distributions of the two percentiles show a significant overlap, which implies that considering mid-level lapse rates alone is not sufficient to infer a high probability of thunderstorm occurrence. 

Complementarily to mid-level lapse rates, we show the distributions of CAPE for the two percentiles in \cref{fig:lapse-rates}c. Most samples in both percentiles have a low value of CAPE, with around \SI{50}{\percent} of the high-probability samples and \SI{90}{\percent} of the low-probability samples falling into the lowest bin. Nevertheless, the top percentile distribution has a longer tail, towards higher CAPE values, than the bottom percentile distribution does. The low CAPE values of the bottom percentile samples are consistent with conditional instability failing to develop. As for the top percentile samples, we expect CAPE to be considerably reduced inside the core of a convective cell, whereas some CAPE is expected to remain in lightning regions of less intense precipitation, producing the long tail of the top percentile distribution.

As pressure saliency is non-vanishing also near the surface, we show in \cref{fig:lapse-rates}b distributions of near-surface (10-\SI{1000}{\meter}) lapse rates. Indeed, the distributions differ for the two percentiles. In contrast to the mid-level peak, near-surface lapse rates for the top percentile tend to be lower, and mostly absolutely stable, whereas lapse rates for the bottom percentile are mostly conditionally unstable. \Cref{fig:lapse-rates}d shows the corresponding distributions of convective inhibition (CIN). While most samples of both percentiles fall into the lowest bin, the top percentile distribution has a longer tail towards higher values of CIN. This might seem surprising as this result seems to suggest that more stable low-level lapse rates (or: higher values of CIN) favor thunderstorm occurrence. However, the reduction of CIN is only conducive to thunderstorm development if sufficient CAPE has been able to build up beforehand. The bottom-percentile samples are essentially from environments with low CAPE and low CIN, which suggests that in these cases, the constantly low CIN (due to, e.g., continuous mixing of low-level air) caused any instabilities aloft to be released prematurely, preventing CAPE from building up, or thunderstorms from forming \citep{Carlson1983,Tuckman2023}. As for the top-percentile samples, CIN is expected to be mostly removed within storm centers, while some CIN can be expected to remain in the less intense regions of convective precipitation or for samples with simulated convection occurring nearby or in the near future. This effect likely produces the longer tail of the top-percentile distribution of CIN.

In summary, SALAMA 1D appears to rely on two categories of patterns. One category consists of patterns related to
\begin{itemize}
    \item tropopausal ice particle content,
    \item vertical wind speed, or
    \item cloud cover.
\end{itemize}
These patterns pertain to regions within ongoing convection, in which precipitation is most intense. In contrast, we identify a category of patterns related to 
\begin{itemize}
    \item horizontal wind direction,
    \item mid-level and near-surface lapse rates, or
    \item low-level moisture.
\end{itemize}
We refer to the latter patterns as mesoscale since the underlying fields, such as temperature and pressure, vary more slowly in the horizontal than the fine-grained hydrometeor variables do. Conversely, we refer to the former patterns as sub-mesoscale.
Mesoscale patterns in the above sense tend to be characteristic of regions with sufficient distance to the convective cores such that precipitation is less intense and some CAPE remains to be released. SALAMA 1D is sensitive to both categories in order to identify thunderstorm occurrence. While sub-mesoscale patterns are useful for the identification of convective storm centers, the sensitivity to mesoscale patterns could explain the ML model's skill at increasing the areas of simulated convection observed in the case study (\cref{fig:case_study}), namely by accounting for modestly-precipitating regions with low-level moisture and left-over CAPE.

\section{Discussion and conclusion}\label{sec:conclusion}
Bypassing the traditional use of derived single-level predictors from NWP data, we developed SALAMA 1D, an ML model for predicting the probability of thunderstorm occurrence on a pixel-wise basis by processing vertical profiles of three-dimensional variables from convection-permitting NWP forecasts. The model's architecture was motivated by physical considerations. In particular, a sparse layer reduces parameter size by encouraging interactions at similar height levels, while a shuffling mechanism prevents the model from learning patterns tied to the vertical grid structure. The latter also adds a form of regularization that limits overfitting.

In comparison to SALAMA 0D, an ML baseline model that infers thunderstorm occurrence from derived single-level features, our model demonstrated higher skill across a wide range of metrics and for lead times up to at least \SI{11}{\hour}. This result indicated that information relevant to thunderstorm occurrence, while intricately encoded in vertical profiles, can be successfully extracted by machine learning, resulting in an improved ability to recognize thunderstorm occurrence in NWP forecasts. Notably, our model remains lightweight in terms of computational complexity, making it just as suitable for real-time operational use as SALAMA 0D is. Case studies suggested that SALAMA 1D is capable of correcting the raw NWP output when convective areas are of incorrect size or when NWP fails to produce convection in the first place. Furthermore, doubling the number of days used to compile the training set (while keeping the training set size constant) also increased skill, underscoring the importance of a large and diverse database of NWP data. We anticipate further skill improvements with the collection of more NWP data.

A sensitivity analysis based on saliency maps revealed that many learned patterns are physically interpretable. For instance, our results suggested that SALAMA 1D has learned the climatological propagation direction of thunderstorms in the study region and relies on fine-grained (sub-mesoscale) structures, such as ice particle content near the tropopause, or cloud cover, to identify high-precipitation regions of ongoing convection. Conversely, mesoscale patterns related to atmospheric instability and moisture are used, possibly to account for regions with less intense precipitation and left-over CAPE. We hypothesized that mesoscale patterns are instrumental in correcting the areal size of simulated convection.

To improve the ML model's capability of correcting for NWP-related biases, it may be beneficial to adapt the model in such a way that horizontally extended input, or several forecast times around the target time, are processed. This would allow for improvement on correcting location and timing errors of convection.
Furthermore, one could train the model in such a way that it can process all ensemble members simultaneously in one forward pass, which would enable the ML model to account for the NWP forecast uncertainty.

In closing, we stress that the methodology applied in this work may be useful for machine learning (ML) topics beyond the identification of thunderstorm occurrence in NWP data. Our approach demonstrates how incorporating physical constraints and symmetry principles can lead to more robust and computationally efficient ML models. Additionally, our use of saliency maps highlights a path toward more interpretable ML models, fostering greater trust as we gain insight into how models arrive at their predictions. As ML continues to play a growing role in severe weather forecasting, ensuring that these models are both accurate and transparent will be key to enhancing their operational utility in critical decision-making processes.

\clearpage
\acknowledgments
We gratefully acknowledge the computational and data resources provided through the joint high-performance data analytics (HPDA) project "terrabyte" of the DLR and the Leibniz Supercomputing Center (LRZ). We thank Björn Brötz and Hessel Juliust for reviewing the ML code. C.M. carried out his contributions within the Italia--Deutschland Science--4--Services Network in Weather and Climate (IDEA-S4S; INVACODA, 4823IDEAP6). This Italian-German research network of universities, research institutes and DWD is funded by the Federal Ministry of Digital and Transport (BMDV). The authors declare that there are no conflicts of interest to disclose.

%
%
\datastatement

The training, validation and testing data sets generated in this study are available on Zenodo (https://doi.org/10.5281/zenodo.13981207). The code for all ML models is released on GitHub (https://github.com/kvahidyou/SALAMA).
The original simulation data from the ICON-D2-EPS model is accessible through the corresponding DWD database (https://www.dwd.de/EN/ourservices/pamore/pamore). Please note that the original data from the LINET lightning detection network is proprietary and cannot be shared in its raw form.

%






%



\bibliographystyle{ametsocV6}
\bibliography{main.bbl}

\end{document}